\documentclass{aa}
\usepackage{aabib}
\usepackage{graphicx}
\begin{document}
\title{...}
   \title{ROSAT-HRI detection of the Class~I protostar YLW16A\\ in the $\rho$ Ophiuchi dark cloud}


   \author{N. Grosso
 	  }

   \offprints{N. Grosso}

   \institute{Max-Planck-Institut f{\"u}r extraterrestrische Physik,
P.O. Box 1312, D-85741 Garching bei M{\"u}nchen, Germany\\
              \email{ngrosso@xray.mpe.mpg.de}
             }

   \date{Received ; accepted }

   \abstract{I analyze unpublished or partially published archival ROSAT data of the $\rho$ Ophiuchi dark cloud. 
This set of seven overlapping ROSAT HRI pointings, composed of eight $\sim$one-hour exposures, 
detects mainly the X-ray brightest T~Tauri stars of this star-forming region. 
Only two HRI sources are new X-ray sources, and their optical counterparts are proposed as new Weak T~Tauri star candidates.
Meanwhile the ROSAT HRI caught during just one exposure a weak X-ray source ($\cal L$=10; 
$SNR$$=$4.1\,$\sigma$ for Gaussian statistics) among a group of three embedded young stellar objects including two Class~I protostars. 
Previous ROSAT PSPC, ASCA GIS observations, and as I argue here one {\it Einstein} IPC observation, 
have already detected an X-ray source in this area, but this higher angular resolution data
show clearly that X-rays are emitted by the Class~I protostar YLW16A. 
This is the second Class~I protostar detected by the ROSAT HRI in this dark cloud.
The determination of the intrinsic X-ray luminosity of this event, $L_\mathrm{X}$[0.1--2.4\,keV]=(9.4--450)$\times 10^{30}$\,erg\,s$^{-1}$, 
critically depends on the source absorption estimate. Improvements will be obtained only by the direct determination 
of this parameter from fitting of Chandra and XMM-Newton spectra. 
               }

   \maketitle
%

\section{Introduction}

As in the 80's the {\it Einstein} X-ray observatory discovered the high X-ray variability 
of low-mass pre-main sequence stars, called T~Tauri stars, the ROSAT satellite 
reported in the 90's the first clues of the X-ray activity of younger stellar objects ($\sim10^5$\,yr), 
called Class~I protostars (\cite{FM99}).
Class~I protostars (\cite{lada91}) are composite objects including a central forming star surrounded by an accretion disk 
$\sim$10--100\,AU in radius, 
and embedded in an extended infalling remnant envelope of gas and dust up to $\sim10^4$\,AU in size (\cite{shu87}).  
Using the {\it Position Sensitive Proportional Counter} (PSPC), \cite*{casanova95} reported 
in the {\object{$\rho$~Ophiuchi star-forming region} ($\sim$145\,pc; \cite{dezeeuw99}) the detection of seven 
X-ray sources coinciding with Class~I protostars, but possible faint non-protostellar counterparts 
were also found in the same error boxes. 
The ASCA satellite detected more clearly a group of five Class~I protostars in the \object{R~CrA} star-forming region ($\sim$130\,pc), 
but the  positional uncertainties was quite large ($\sim20\arcsec$), and the individual source only 
partially resolved (\cite{koyama96}). In both cases ROSAT, with its {\it High Resolution Imager} (HRI) providing a much better 
spatial resolution (FWHM$\approx$5\arcsec), has helped to confirm these detections 
(\cite{grosso97}, hereafter paper~I; \cite{neuhaeuser97}).  
With the goal of preparing the upcoming study of these two star-forming regions with the new generation of X-ray 
satellites, Chandra and XMM-Newton, their ROSAT observations must be definitively exploited. 
This work was already done by \cite*{neuhaeuser97} on R~CrA. 
I study here the unpublished or partially published observations of the $\rho$ Ophiuchi dark cloud 
to search for X-rays from Class~I protostars.

\begin{table*}[!ht]
\caption[]{List of the ROSAT observations of the $\rho$ Oph dark cloud. Only the segments \#1--8 are analyzed here.}
\label{tab:rosat_list}
\begin{tabular}{lccrcccclc}
\hline
\noalign{\smallskip}
\multicolumn{1}{c}{ROSAT} 	  & Proposal name       & PI name        & \multicolumn{1}{c}{Exp.}      & Start    & End      
& $\alpha_{\mathrm{J2000}}$ & $\delta_{\mathrm{J2000}}$    & Ref. & Segment\\
 \multicolumn{1}{c}{archive}        &                     &                & \multicolumn{1}{c}{[ks]}       
& [yymmdd] & [yymmdd] & 16$^{\mathrm h}$	& -24$^\circ$	&  & \#	\\
\noalign{\smallskip}
\hline
\noalign{\smallskip}
200045p-0 & $\rho$~Oph Core     &    Montmerle   &    12.9 &   910305 & 913010 &  26$^{\mathrm m}$31$\fs$0 & 31$^\prime$48$^{\prime\prime}$ & (1)	&\\      
200045p-1 & $\rho$~Oph Core     &    Montmerle   &    19.9 &   910908 & 910908 &  26$^{\mathrm m}$31$\fs$0 & 31$^\prime$48$^{\prime\prime}$ & (1)	&\\  
201709h   &  SR9, SR12          &    Damiani     &     3.3 &   940829 & 940829 &  27$^{\mathrm m}$28$\fs$1 & 31$^\prime$48$^{\prime\prime}$ & (2)	& 1 \\        
201710h   &  SR9, SR12          &    Damiani     &     5.2 &   940831 & 940831 &  27$^{\mathrm m}$28$\fs$1 & 31$^\prime$48$^{\prime\prime}$ & (2)	& 2 \\      
201711h   &  SR9, SR12          &    Damiani     &     2.7 &   940901 & 940901 &  27$^{\mathrm m}$28$\fs$1 & 31$^\prime$48$^{\prime\prime}$ & (2)	& 3 \\       
201712h   &  SR9, SR12          &    Damiani     &     5.1 &   940903 & 940903 &  27$^{\mathrm m}$28$\fs$1 & 31$^\prime$48$^{\prime\prime}$ & (2)	& 4 \\       
201713h   &  SR9, SR12          &    Damiani     &     8.1 &   940904 & 940904 &  27$^{\mathrm m}$28$\fs$1 & 31$^\prime$48$^{\prime\prime}$ & (2)	& 5 \\        
{\bf 201714h}   &  SR9, SR12  &    Damiani     &     {\bf 3.7} &   {\bf 940905} & {\bf 940905} &  27$^{\mathrm m}$28$\fs$1 & 31$^\prime$48$^{\prime\prime}$ & (2)& {\bf 6} \\ 
201618h-1 &  ROX20              &    Zinnecker   &     1.9 &   940917 & 940917 &  27$^{\mathrm m}$14$\fs$0 & 51$^\prime$36$^{\prime\prime}$ &   	& 7 \\    
201834h   &  ~Oph Core F        &    Montmerle   &    12.6 &   950309 & 950314 &  27$^{\mathrm m}$26$\fs$0 & 40$^\prime$48$^{\prime\prime}$ & (3, 4)	&   \\     
201618h-2 &  ROX20              &    Zinnecker   &     3.6 &   950817 & 950817 &  27$^{\mathrm m}$14$\fs$0 & 51$^\prime$36$^{\prime\prime}$ & 		& 8 \\      
201834h-1 &  $\rho$~Oph Core F  &    Montmerle   &    27.8 &   950818 & 950820 &  27$^{\mathrm m}$26$\fs$0 & 40$^\prime$48$^{\prime\prime}$ & (3, 4)	&\\      
201835h   &  $\rho$~Oph Core A  &    Montmerle   &    51.7 &   950829 & 950912 &  26$^{\mathrm m}$02$\fs$0 & 23$^\prime$24$^{\prime\prime}$ & (4)	&\\
201834h-2 &  $\rho$~Oph Core F  &    Montmerle   &    37.5 &   960907 & 960911 &  27$^{\mathrm m}$26$\fs$0 & 40$^\prime$48$^{\prime\prime}$ & (4)	&\\    
\noalign{\smallskip}
\hline   
\noalign{\smallskip}
\end{tabular}\\
Note: In the ROSAT archive ``p'' (resp. ``h'') is for the PSPC (resp. HRI) instrument.\\
References: (1) \cite*{casanova95}; (2) \cite*{damiani96}; (3) paper~I; (4) \cite*{grosso00}.
\end{table*}

\section{ROSAT observations}

I found in the ROSAT archive that ten overlapping pointings were performed on the $\rho$ Ophiuchi dark cloud: 
one with the PSPC and nine with the HRI (see Table~\ref{tab:rosat_list} for the log of these observations).
Among these HRI pointings a sequence of six short exposures ($\sim$ hour) spreaded over one week, 
was used by \cite*{damiani96} to study the X-ray variability of the T Tauri stars SR9 and SR12A-B.
One more HRI pointing remains without published results.
I thus analyzed these data (8 segments in all; see Table~\ref{tab:rosat_list}) with EXSAS (\cite{zimmermann97}).

Source detection was performed on each segment separately with the standard command {\tt DETECT/SOURCES}, 
which generates a local source detection by a sliding-window technique followed by a maximum likelihood test, 
which compares the observed count distribution to a model of the point spread function and the local 
background to discriminate sources from statistical Poissonian background fluctuations.  
The {\it likelihood of existence}, defined as ${\cal L}$=$-\ln P_0$ (with $P_0$ the probability 
of the null hypothesis that the observed distribution of counts is only due to a statistical 
background fluctuation), provides a maximum likelihood measure for the source detection. 
I accepted only detections with ${\cal L}$$\ge$10 ($SNR$$\ge$4.1\,$\sigma$ for Gaussian statistics) 
to reduce the number of spurious detections per field to $\sim$0.7--$\sim$1.0.
Identification of these X-ray sources was made by cross-correlation with published list of confirmed 
or suspected cloud members (e.g. \cite{am94}), IR surveys (e.g. \cite{bklt97}), and optical catalogue (\cite{monet96}).
To correct X-ray positions for boresight errors, I selected the X-ray sources not associated with protostars 
having a positional error lower than $2\arcsec$ both in $\alpha$ and $\delta$, and compared their positions with 
their IR counterparts in the 2MASS catalogue (second incremental release; \cite{cutri00});
the mean offsets in $\alpha$ and $\delta$ was then substracted to the X-ray positions of all X-ray sources, 
and the residual dispersion was quadratically added to the positional errors.  
Table~\ref{tab:rosat_sources} gives the X-ray source list.

    \begin{figure}[!ht]
   \centering
   \includegraphics[width=\columnwidth]{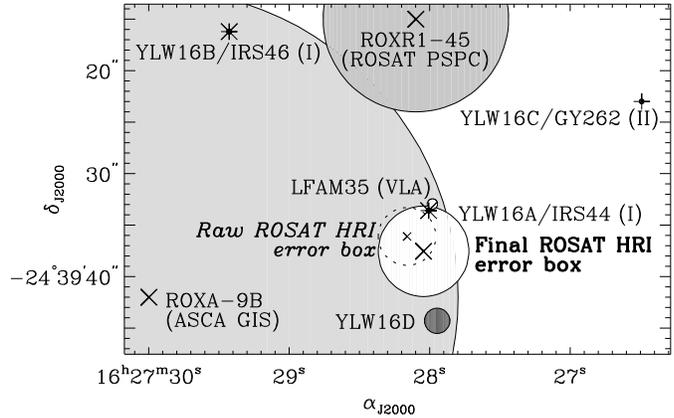}
      \caption{X-ray detection of the Class~I protostar YLW16A with the ROSAT HRI. 
The positions of the IR sources in this area are from the 2MASS catalogue (Cutri et al. 2000; $\sigma_\mathrm{2MASS}\sim0.3\arcsec$). 
The position of YLW16D, given in the Barsony et al. (1997) coordinate frame by Lucas \& Roche (1998), 
was converted to the 2MASS coordinate frame applying the position offset found for YLW16A in 2MASS compared to Barsony et al. (1997). 
The position of LFAM35, the VLA detection of YLW16A (Leous et al. 1991), is also shown. 
The circle radii indicate with the one $\sigma$ positional error of the different observations.  
The ROSAT HRI error circle after boresight error correction includes only YLW16A. 
              }
         \label{astrometry}
   \end{figure}

\begin{table}[!ht]
\caption[]{X-ray sources detected by the ROSAT HRI with a likelihood of existence $\ge$10 in the segments \#1--8. 
Col.~(1) gives the IR/optical counterpart name with :
GY = Greene \& Young (1989);
IRS = Wilking et~al. (1989); 
ROXs = Bouvier \& Appenzeller (1992); 
SR = Struve \& Rudkj{\o}bing (1949); 
YLW = Young et~al. (1986); 
numbers are from the PMM USNO-A1.0 catalogue (Monet et al. 1996; the designation prefix {\it 0600-1} was here cut away).
Col. (2) indicates the IR classification.
The better X-ray positions of this segment set in Col.~(3--5) are corrected from boresight errors.
The segment number as defined in Table~\ref{tab:rosat_list} is given in Col.~(6).
$\cal{L}$, in Col.~(7), is the likelihood of existence of the X-ray sources for each observation.
The count rates are given in the ROSAT 0.1--2.4\,keV energy band in Col.~(8).}
\label{tab:rosat_sources}
\resizebox*{\columnwidth}{!}{
\begin{tabular}{@{}lllrrr@{}}
\hline
\noalign{\smallskip}
\multicolumn{1}{c}{Source} & \multicolumn{1}{c}{IR} & ~$\alpha_\mathrm{J2000}$~~~~~$\delta_\mathrm{J2000}$~~~~~\,$\pm$ 	& \multicolumn{1}{c}{\#} & \multicolumn{1}{c}{$\cal{L}$} 	& \multicolumn{1}{c}{count rate} \\
\multicolumn{1}{c}{Name}   & \multicolumn{1}{c}{Cl.} &  ~~~16$^\mathrm{h}$ ~~~~~~~~~~~~~~~~~~[$\arcsec$]	& &		& \multicolumn{1}{c}{[cts\,ks$^{-1}$]} \\
\multicolumn{1}{c}{(1)} & \multicolumn{1}{c}{(2)} & ~~~(3)~~~~~~~~~(4)~~~~~~\,(5) & \multicolumn{1}{c}{(6)} & \multicolumn{1}{c}{(7)} & \multicolumn{1}{c}{(8)} \\
\noalign{\smallskip}
\hline
\noalign{\smallskip}
     \object{1589441} & III? &        26$^\mathrm{m}$04$\fs$5~-24\degr57\arcmin50\arcsec~~5 &         8 &    54.1 & $20.2\pm 3.0$ \\
       \object{SR24N} &   II &        26$^\mathrm{m}$58$\fs$3~-24\degr45\arcmin32\arcsec~~3 &         7 &    10.8 & $ 4.6\pm 1.7$ \\
                     &      &                                                              &         8 &    13.3 & $ 2.4\pm 0.9$ \\
       \object{GY194} &  III &        27$^\mathrm{m}$04$\fs$8~-24\degr42\arcmin18\arcsec~~5 &         7 &    10.1 & $ 3.9\pm 1.6$ \\
     \object{ROXs20B} &  III &        27$^\mathrm{m}$15$\fs$1~-24\degr51\arcmin40\arcsec~~3 &         7 &    18.5 & $ 4.0\pm 1.5$ \\
                     &      &                                                              &         8 &    38.0 & $ 4.5\pm 1.2$ \\
     \object{SR12A-B} &  III &        27$^\mathrm{m}$19$\fs$7~-24\degr41\arcmin39\arcsec~~2 &         1 &   143.2 & $16.7\pm 2.4$ \\
                     &      &                                                              &         2 &   123.0 & $11.3\pm 1.6$ \\
                     &      &                                                              &         3 &   138.6 & $24.2\pm 3.2$ \\
                     &      &                                                              &         4 &   128.2 & $12.3\pm 1.7$ \\
                     &      &                                                              &         5 &  2150.2 & $85.4\pm 3.4$ \\
                     &      &                                                              &         6 &    90.5 & $13.0\pm 2.0$ \\
                     &      &                                                              &         7 &    42.9 & $11.7\pm 2.7$ \\
                     &      &                                                              &         8 &   162.3 & $19.0\pm 2.4$ \\
      \object{YLW16A} &    I &        27$^\mathrm{m}$28$\fs$0~-24\degr39\arcmin38\arcsec~~4 &         6 &    10.0 & $ 2.2\pm 0.9$ \\
     \object{1614157} & III? &        27$^\mathrm{m}$32$\fs$5~-25\degr06\arcmin16\arcsec~~6 &         7 &    21.3 & $12.1\pm 3.1$ \\
                     &      &                                                              &         8 &    20.8 & $ 8.2\pm 1.9$ \\
       \object{GY292} &   II &        27$^\mathrm{m}$33$\fs$1~-24\degr41\arcmin20\arcsec~~5 &         2 &    11.8 & $ 2.3\pm 0.8$ \\
                     &      &                                                              &         4 &    10.7 & $ 2.0\pm 0.7$ \\
       \object{IRS49} &   II &        27$^\mathrm{m}$38$\fs$2~-24\degr36\arcmin58\arcsec~~3 &         6 &    10.3 & $ 1.8\pm 0.8$ \\
         \object{SR9} &   II &        27$^\mathrm{m}$40$\fs$4~-24\degr22\arcmin01\arcsec~~3 &         1 &   244.9 & $29.5\pm 3.2$ \\
                     &      &                                                              &         2 &   582.6 & $38.7\pm 2.9$ \\
                     &      &                                                              &         3 &   237.5 & $36.7\pm 3.9$ \\
                     &      &                                                              &         4 &   572.2 & $38.4\pm 2.9$ \\
                     &      &                                                              &         5 &  1012.7 & $47.5\pm 2.6$ \\
                     &      &                                                              &         6 &   367.0 & $34.8\pm 3.2$ \\
      \object{ROXs31} &  III &        27$^\mathrm{m}$52$\fs$2~-24\degr40\arcmin53\arcsec~~3 &         1 &    19.5 & $ 4.8\pm 1.4$ \\
                     &      &                                                              &         5 &    29.8 & $ 4.6\pm 0.9$ \\
                     &      &                                                              &         6 &    42.1 & $ 7.4\pm 1.6$ \\
        \object{SR20} &  III &        28$^\mathrm{m}$33$\fs$0~-24\degr22\arcmin55\arcsec~~9 &         1 &    12.0 & $ 7.2\pm 2.1$ \\
        \object{SR13} &   II &        28$^\mathrm{m}$45$\fs$5~-24\degr28\arcmin22\arcsec~~5 &         1 &    33.7 & $14.4\pm 2.7$ \\
                     &      &                                                              &         2 &    18.9 & $10.9\pm 2.2$ \\
                     &      &                                                              &         3 &    11.2 & $10.7\pm 3.0$ \\
                     &      &                                                              &         4 &    96.2 & $21.5\pm 2.5$ \\
                     &      &                                                              &         5 &    66.4 & $16.6\pm 2.0$ \\
                     &      &                                                              &         6 &    26.2 & $12.3\pm 2.5$ \\
\noalign{\smallskip}
\hline
\end{tabular}}
\end{table}

   \begin{figure}[!ht]
   \centering
   \includegraphics[width=\columnwidth]{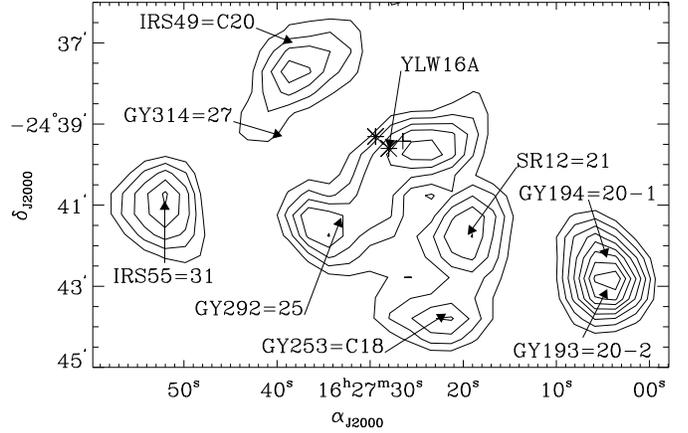}
      \caption{{\it Einstein} IPC detection of the YLW16 IR source group. This intensity contour map is an enlargement 
of a convolution of the counts distribution (pixel size binned to $30\arcsec$) of an archival 
{\it Einstein} IPC observation  of the $\rho$ Ophiuchi dark cloud (8 sep. 1979; 2.1\,ks exposure) with a Gaussian profile 
(FWHM=$1\arcmin$). 
The image background is $B$=2.5$\times 10^{-4}$\,cts\,s$^{-1}$\,arcmin$^{-2}$.
The intensity scale is linear, and was choosen so that the first and fourth contours match the Montmerle et al. 
(1983)'s contours $n$=1 and 4, and thus represent levels $2\times1.5^{1+n}B$ above standard background.
The {\it Einstein} IPC FWHM in this area, located $\sim20\arcmin$ away from the axis, is larger than $1\arcmin$.   
The arrow heads point the 2MASS positions of the IR counterpart of {\it Einstein} IPC sources 
(the ROX number is given after the sign ``=''). 
Asterisks (resp. cross) show the Class~I protostar (resp. T~Tauri star) positions (see also Fig.~\ref{astrometry}). 
An {\it Einstein} X-ray source is associated with the YLW16 IR source group.}
         \label{ipc}
   \end{figure}

These short exposures detected mainly the X-ray brightest T~Tauri stars of this star-forming region. 
Only two HRI sources are new X-ray sources. They are located $\sim$25$\arcmin$ south ($\sim$1\,pc)
of the core E/F (\cite{loren90}) and identified with optical stars\footnote{11589441 ($B$=14.8, $R$=12.6); 
 11614157 ($B$=17.9, $R$=14.1).}: they are reliable Weak T~Tauri star candidates.
Meanwhile a weak HRI source (${\cal L}$=10), detected just in segment 8, 
is associated with the IRAS source YLW16 -- a group of two Class~I protostars (YLW16A; and YLW16B, 
also called IRS46) and one embedded Classical T~Tauri star (GY262).
In order to discuss the possible identification of this weak HRI source with protostars, 
Fig.~\ref{astrometry} visualizes the different members of YLW16 in the 2MASS reference frame, 
to be consistent with my previous astrometric correction. 
The 2MASS position of YLW16A is within the $4\arcsec$-radius error circle of the HRI source.
Using a Monte-Carlo simulation, I found that the probability that this identification is only due to chance is $1.6\%$.
This reliable identification strenghtens the significance level of this weak HRI detection.

As shown in Fig.~\ref{astrometry}, other instruments detected also an X-ray source in this area. 
The ROSAT PSPC source ROXR1-45 (\cite{casanova95}) was identified with YLW16. 
The ASCA GIS source ROXA-9B (\cite{kamata97}) was only identified with YLW16B, 
whereas YLW16A is also in the ASCA GIS positional error box as mentioned by \cite*{carkner98}; 
moreover the new IR source discovered by \cite*{lucas98}, YLW16D, is in the same error box.
An ASCA follow-up observation, made 3.5 years after, did not detect again ROXA-9B (Tsuboi et al. 1997): 
this X-ray source is variable.
I noted also in only one of the {\it Einstein} IPC observations of Montmerle et al. 
(1983; see observation I3749.2 Fig.~1.2) an X-ray emission around YLW16, 
blended with the {\it Einstein} X-ray sources associated with SR12 
(ROX21) and GY292 (ROX25), which was not in the Montmerle et al.'s detection list.
To solve this point, I took from HEASARC the archived screened photon event 
list\footnote{Sum of I3749.1 (8 March 1979), and I3749.2 (8 Sep. 1979).}, 
and revisited it using the package XANADU/XIMAGE (also available at HEASARC). 
I selected only the 8 Sep. 1979 events and energy bins 3--13 (corresponding to 0.3--8.2\,keV). 
From this event list, I studied the X-ray source standing near YLW16 with the interactive command 
{\tt sosta}, which allows to estimate the background from a nearby area free of sources, 
and to tune the source box size to exclude events from neighbouring sources.    
I found for this source a signal to noise ratio $\sim$3.5\,$\sigma$, 
and a corrected intensity $S$$\sim$0.02\,cts\,s$^{-1}$. 
Fig.~\ref{ipc} shows a contour map constructed from these data. 
I conclude that the {\it Einstein} IPC also detected X-rays from YLW16 
(which was not yet discovered from IR at this time), 
but the {\it Einstein} IPC FWHM ($\sim$1\arcmin) is too large to find unambiguously the IR counterpart. 
By contrast to these previous detections with other X-ray instruments, we can see clearly in Fig.~\ref{astrometry}, 
thanks to the better angular resolution of the HRI, that the X-ray emission detected by the ROSAT HRI 
comes only from the Class~I protostar YLW16A. 

\section{X-ray luminosity of YLW16A}

YLW16A is the second Class~I protostar detected by the HRI in the $\rho$ Ophiuchi dark cloud 
after \object{YLW15} (paper~I).
It is remarkable to detect it with only an one-hour HRI exposure, 
whereas no detection was obtained with a 20 times longer HRI exposure (see Table~\ref{tab:rosat_list}; \cite{grosso00}). 
The fact that YLW16A was not detected by the HRI only one day before, with an exposure 2 times longer, implies 
a variation of its X-ray luminosity by at least a factor 2 in less than one day.
This suggests that YLW16A was detected during a state of higher X-ray luminosity, 
probably due to an X-ray flare, usual in young stellar objects (see Feigelson \& Montmerle 1999). 
However the source X-ray variability cannot be tested (e.g. with the Kolmogorov-Smirnov test) with only 
$\sim8$\,cts detected during this short exposure.
It is impossible to deduce from this HRI detection whether the previous X-ray observations by other instruments 
actually detected YLW16A, as this X-ray source is not constant.

The determination of the intrinsic X-ray luminosities of such embedded young stellar object critically depends 
on the absorption of the X-ray photons by the gas along the line of sight, $N_\mathrm{H}$, 
combination of both the circumstellar material and the interstellar medium. 
As the HRI has no spectral resolution, no direct information on $N_\mathrm{H}$ 
can be provided by the X-ray event.
I thus derive $N_\mathrm{H}$ from the source visual extinction due to the dust, $A_\mathrm{V}$, 
assuming the conversion factor $N_\mathrm{H}$=2.23$\times$10$^{21}A_\mathrm{V}$\,mag\,cm$^{-2}$ (\cite{ryter96}).
Applying the methods used in paper~I, I find from near-IR: $A_\mathrm{V}$=30--40, 
and thus $N_\mathrm{H}$=(6.7--8.9)$\times 10^{22}$\,cm$^{-2}$.
For comparison if I assume that ASCA observed the same source, I have directly from the ASCA spectrum 
(the only one with enough statistics): 
$N_\mathrm{H}$=2.8$\times10^{22}$\,cm$^{-2}$, corresponding to $A_\mathrm{V}$=13. 
Such extinction discrepancies between near-IR and X-ray estimates are not unusual for Class~I protostars, 
but are not well understood (see \cite{kamata97}), and thus this low value of $N_\mathrm{H}$ cannot be excluded.
Taking the range $N_\mathrm{H}$=(2.8--8.9)$\times 10^{22}$\,cm$^{-2}$ for the absorption, a distance of 145\,pc, 
an isothermal Raymond-Smith plasma spectrum with a typical protostar flare temperature $kT$=4\,keV, 
standard solar elemental abundances, I find using {\tt W3PIMMS} 
(see HEASARC homepage): $L_\mathrm{X}$[0.1--2.4\,keV]=(9.4--450)$\times10^{30}$\,erg\,s$^{-1}$ 
(for comparison $L_\mathrm{bol}$$\sim$13$\,L_\odot$; \cite{wilking89}), i.e. a factor $\sim50$ of uncertainties remains in the X-ray 
luminosity of this events.
Nevertheless, this X-ray luminosity is comparable to the one observed during the X-ray triple flare detected from YLW15 by ASCA (\cite{tsuboi00}). 

Improvements of such studies will come from the Chandra and XMM-Newton observations, 
which will give unambiguous X-ray spectra of YLW16A, 
and thus will provide an accurate value of $N_\mathrm{H}$ 
to constrain the X-ray luminosity observed by the HRI.

\begin{acknowledgements}
I thank the anonymous referee for valuable comments, 
and T. Montmerle for discussions about his pioneering work with 
the {\it Einstein} observatory.
I would like also to thank my host institution, 
and particularly Prof. J. Tr{\"u}mper and R. Neuh{\"a}user, 
who have welcomed me for my Marie Curie Individual fellowship 
supported by the European Union (HPMF-CT-1999-00228). 

\end{acknowledgements}

\end{document}